\makeatletter
\def\input@path{{styles/IEEE/}}
\makeatother
\documentclass[lettersize,journal]{IEEEtran}

\usepackage{cite}
\usepackage{amsmath,amsfonts}
\usepackage{graphicx}
\usepackage{textcomp}
\usepackage{xcolor}
\usepackage{xspace}
\usepackage{verbatim}
\usepackage{algorithm}
\usepackage{algpseudocode,algorithmicx}
\usepackage{makecell}
\usepackage{hyperref}
\usepackage{multirow}
\usepackage{array}
\usepackage{booktabs}
\usepackage{placeins}
\usepackage[caption=false,font=footnotesize]{subfig}

\usepackage{microtype}

\usepackage{eso-pic}
\AddToShipoutPictureFG*{%
  \AtPageLowerLeft{%
    \put(\LenToUnit{0.5\paperwidth},\LenToUnit{0.22in}){%
      \makebox[0pt]{\parbox{0.9\paperwidth}{\centering\scriptsize
        \copyright~2026 IEEE. Personal use of this material is permitted. Permission from IEEE must be obtained for all other uses, in any current or future media, including reprinting/republishing this material for advertising or promotional purposes, creating new collective works, for resale or redistribution to servers or lists, or reuse of any copyrighted component of this work in other works.\\
        Accepted for publication in IEEE Computer Architecture Letters.}}}}}

\newcommand{\name}[0]{{LogPIM}\xspace}

\begin{document}

\title{\name: Low-Overhead Logarithmic Arithmetic for HBM-Based Processing-in-Memory}

\author{Taewoon Kang, Jonghyun Lee, Yujin Kim, Murali Annavaram, and Gunjae Koo

\thanks{

Taewoon Kang, Yujin Kim, and Gunjae Koo are with the Department of Computer Science and Engineering, Korea University, Seoul, 02841, South Korea (e-mail: taewoon\_kang@korea.ac.kr; yujin\_kim@korea.ac.kr; gunjaekoo@korea.ac.kr). 
Jonghyun Lee, and Murali Annavaram are with the Ming-Hsieh Department of Electrical Engineering, University of Southern California, Los Angeles, CA 90089, USA (e-mail: leejongh@usc.edu; annavara@usc.edu).
Gunjae Koo is the corresponding author. }}

\maketitle

\begin{abstract} 
Processing-in-memory (PIM) reduces data movement cost, but integrating processing units into DRAM memory cells incurs substantial area overhead and reduces available memory capacity. We propose \name, an LNS-based PIM architecture that replaces FP16 multipliers with dual-mode SIMD adders, performing LNS multiplication using fixed-point addition and FP16 accumulation. The proposed LUT controller performs LNS-to-FP16 conversion using one DRAM row per bank, while the ping-pong mechanism overlaps matrix and LUT accesses across paired banks to avoid repeated row switching. Compared to conventional HBM-PIM, \name reduces processing-unit area and GeMV kernel energy by 30.1\% and 33.9\%, respectively. \name achieves a $1.68\times$ speedup on average for PIM kernels with pre-encoded LNS16 inputs. 
The LNS16-based GeMV outputs by \name achieve an average cosine similarity of 99.99\% to the FP32 reference.
Our evaluation demonstrates that \name improves area, energy, and performance for HBM-based PIM.
\end{abstract}


\section{Introduction}\label{sec:intro}

\begin{figure}
    \centerline{\includegraphics[width=0.60\columnwidth]{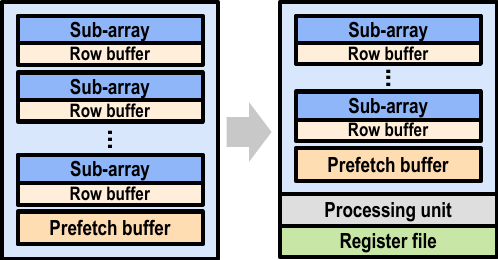}}
    \vspace{-3.5mm}
    \caption{HBM-based PIM sub-array loss}
    \label{fig:hbm_pim_capacity_loss}
    \vspace{-5mm}
\end{figure}

Processing-in-memory (PIM) reduces data movement and improves bandwidth utilization by offloading computation into memory. However, high-bandwidth memory (HBM)-based PIM replaces DRAM sub-arrays with processing units, reducing the capacity available for applications and models (Fig.~\ref{fig:hbm_pim_capacity_loss})~\cite{hbm_pim}. Thus, reducing the area and power overhead by processing units in PIM architectures is critical for mitigating the computation--capacity trade-off.

One of the promising approaches to addressing the aforementioned challenge is to simplify the arithmetic units embedded inside memory. Among various arithmetic representations, the logarithmic number system (LNS) has been recognized as an attractive alternative to conventional floating-point arithmetic since it can fundamentally reduce arithmetic complexity~\cite{logflex,lns-llm}. Unlike conventional floating-point arithmetic, LNS represents numbers in the logarithmic domain, turning multiplication and division into simple addition and subtraction. Consequently, complex floating-point multipliers can be replaced with simpler arithmetic hardware, offering substantial reductions in area and power consumption.

In LNS, a real-valued number is represented as:
\[
X \rightarrow
\begin{cases}
x=\log_b|X|,\\[4pt]
s=
\begin{cases}
0,&\text{if }X>0,\\
1,&\text{if }X<0,
\end{cases}
\end{cases}
\]
where \(x\) denotes the logarithmic magnitude and \(s\) denotes the sign bit. Since zero has no finite logarithmic representation, \name assigns it a reserved codeword in the logarithmic field. For nonzero operands, arithmetic operations in the logarithmic domain are expressed as follows:
\[
\begin{aligned}
\log_b(x \pm y) &= \log_b x + \log_b\!\left(1 \pm b^{\log_b y - \log_b x}\right),\\
\log_b(x \times y) &= \log_b x + \log_b y,\\
\log_b(x \div y) &= \log_b x - \log_b y.
\end{aligned}
\]

The primary limitation of LNS is that addition and subtraction require evaluating nonlinear functions, which are typically implemented using lookup tables (LUTs). Although LUT access cost and area overhead have limited the adoption of LNS in conventional processors, LNS arithmetic is particularly well suited to PIM architectures. Since PIM inherently integrates memory arrays with computation units, LUTs can be stored inside DRAM cells with negligible storage overhead. Thus, PIM provides a promising platform for exploiting the computational advantages of LNS while efficiently supporting LUT-assisted arithmetic.

We propose \textbf{\name}, a LNS-based PIM architecture that simplifies conventional PIM processing units. \name replaces complex FP16 arithmetic units with dual-mode SIMD adders and lightweight LUT controllers. This design substantially reduces area and power overhead, mitigating the memory capacity loss of conventional HBM-PIM architectures while maintaining high-performance in-memory computation.

\section{System overview}\label{sec:overview}
\subsection{Proposed architecture}\label{subsec:proposed_architecture}

\begin{figure}
    \centerline{\includegraphics[width=0.8\columnwidth]{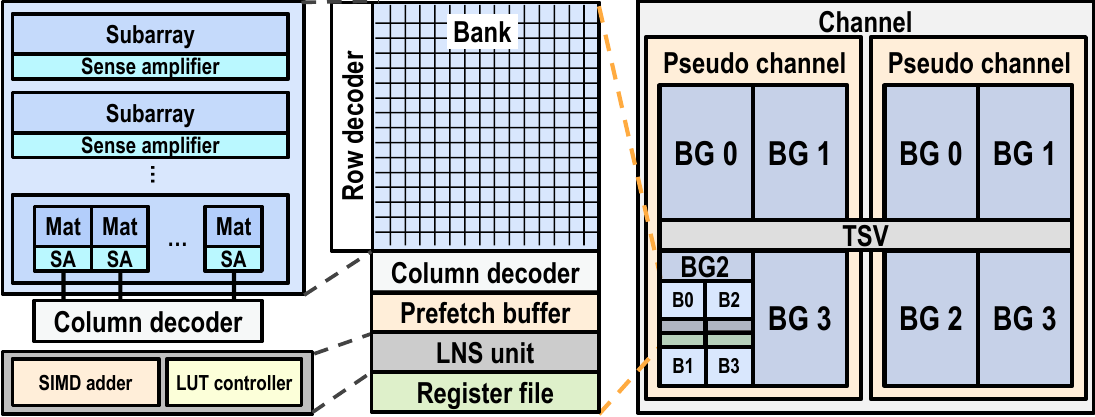}}
    \vspace{-3.5mm}
    \caption{Overall architecture}
    \vspace{-5.5mm}
    \label{fig:overall_architecture}
\end{figure}

Fig.~\ref{fig:overall_architecture} presents the overall architecture of \name. To reduce the hardware overhead of conventional HBM-PIM processing units, \name eliminates the dedicated SIMD FP16 multiplier, redesigns the conventional SIMD FP16 adder as a dual-mode SIMD adder, and introduces a lightweight LUT controller. The proposed SIMD adder supports both FP16 addition and 16-bit fixed-point addition. Since it processes only one numerical format at a time, both modes share as much of the arithmetic datapath as possible, including the significand or fixed-point addition logic, normalization-related components, and register interfaces. The dual-mode SIMD adder includes a small combinational logic that
selects between FP16 and fixed-point operation, without introducing an additional cycle compared to the baseline SIMD multiplication. This hardware reuse reduces the additional area and power required for supporting both logarithmic-domain multiplication and FP16 accumulation. For the register files, \name retains the same configuration as the baseline HBM-PIM architecture~\cite{hbm_pim}, requiring no additional register storage or modification to the existing register organization.

The LUT controller converts LNS products into FP16 values. For each generated LNS product, the controller extracts the LUT index, derives the corresponding DRAM column address within the LUT row, and tracks the addresses of pending requests. To avoid introducing a dedicated tracking buffer, the controller stores these addresses in the unused portion of the existing SRF (scalar register file), requiring no additional register-file capacity. Through tracking, the LUT controller groups requests targeting the same DRAM column and serves them through fine-grained column accesses to the LUT row. Because multiple conversion requests can share the same row activation and, when their LUT entries reside in the same DRAM column, the controller amortizes DRAM access overhead while avoiding a dedicated on-chip LUT structure.

\name adopts a 16-bit base-2 LNS format comprising a 1-bit sign field for \(X\) and a 15-bit signed fixed-point field for \(\ell=\log_2|X|\). The logarithmic field uses a two's-complement I6F9 representation, in which the six integer bits include the sign bit of \(\ell\), and the remaining nine bits represent its fractional component. The 9-bit fractional field indexes 512 16-bit entries storing \(2^F\), yielding a 1-KB conversion LUT that fits in one DRAM row. During input encoding, zero and magnitudes below the minimum nonzero magnitude representable by LNS16 are mapped to the reserved zero codeword. During LNS multiplication, a zero operand directly produces a zero product; otherwise, an underflowed logarithmic sum is mapped to zero, whereas an overflowed sum is saturated to the maximum representable I6F9 value.

For a nonzero product within the representable I6F9 range, the logarithmic magnitude is decomposed as \(\ell=I+F\), where \(I\) denotes the signed integer component and \(F\in[0,1)\) denotes the fractional component. The LUT is indexed by \(F\) to obtain the fractional anti-logarithm \(2^F\), while \(I\) scales the LUT output through a left or right shift. The reconstructed magnitude is then bounded to the supported FP16 range, and the value-sign bit is applied to generate the signed result. This decomposition supports a wide LNS dynamic range while limiting the LUT storage requirement to one DRAM row per bank. 

At the system boundary, we assume a heterogeneous host--PIM execution flow in which a host accelerator, such as a GPU or NPU, converts each FP16 input feature vector to LNS16 before dispatching it to \name. Model weights are converted to LNS16 offline and stored in HBM in the encoded format. Within \name, each LNS product is converted to FP16 before accumulation; therefore, the completed GeMV output is already in FP16 and can be fetched directly by the host accelerator for subsequent operations, such as softmax and reduction.

\subsection{Tiling} \label{subsec:tilig}

\begin{figure}
    \centering
    \vspace{-3mm}
    \subfloat[Previous work\label{fig:tiling-aespa}]{
        \includegraphics[width=0.37\columnwidth]
        {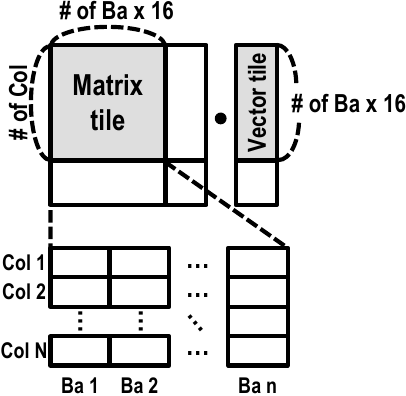}}
    \subfloat[\name\label{fig:tiling-name}]{
        \includegraphics[width=0.25\columnwidth]
        {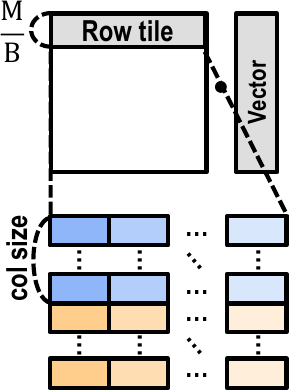}}
    \vspace{-2mm}
    \caption{Comparison of tiling methods for outer-product GeMV}
    \label{fig:tiling}
    \vspace{-5.5mm}
\end{figure}

Fig.~\ref{fig:tiling} compares the GeMV data-placement schemes. While previous outer-product PIMs like AESPA~\cite{aespa} partition matrices along the reduction dimension (Fig.~\ref{fig:tiling}a) and require a bank-group-level shared accumulator to merge partial sums, \name fundamentally eliminates this overhead through an output-row partitioned tiling scheme (Fig.~\ref{fig:tiling-name}).

Given that matrix row sizes are typically multiples of the bank count (e.g., $B=128$), we assign each bank a disjoint set of output rows ($\mathcal{R}_{b}$) and store the corresponding matrix data in a column-wise layout mapped to DRAM columns. Upon receiving a broadcast vector element, each bank performs element-wise multiplication with its local column data and accumulates the products directly into its local registers. Because every term required for a specific output row resides entirely within its assigned bank, each bank independently produces a complete portion of the final result vector. This strictly local accumulation completely removes the necessity for cross-bank reduction and shared accumulators.

\subsection{Basic Execution Flow}\label{subsec:execution_flow}

\begin{figure}
    \centerline{\includegraphics[width=\columnwidth]{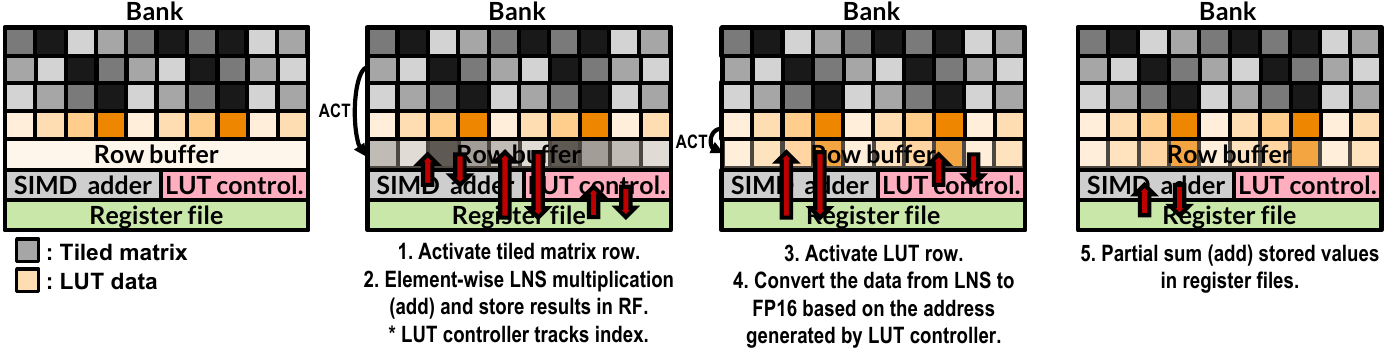}}
    \vspace{-4mm}
    \caption{Basic execution flow}
    \vspace{-5mm}
    \label{fig:execution_flow}
\end{figure}

Fig.~\ref{fig:execution_flow} illustrates the basic per-bank execution flow.%
Each bank reserves one DRAM row for the LUT used for LNS-to-FP16 conversion. The LUT row is stored alongside the tiled matrix rows and accessed through the existing row buffer and sense amplifiers, requiring no dedicated LUT storage structure and incurring only one-row capacity overhead per bank.

At outer-product iteration \(j\), a tiled matrix row is first activated, and the broadcast vector element \(x_j\) is multiplied by the corresponding matrix elements. Since both operands are represented in LNS, each multiplication is reduced to a fixed-point addition performed by the SIMD adder. The register file temporarily stores the resulting LNS products, while the LUT controller generates and tracks the corresponding LUT indices. The bank then activates the reserved LUT row and converts the buffered LNS products to FP16. Multiple conversion requests can be grouped so that one LUT-row activation is shared across several LUT column accesses. The converted products are subsequently accumulated in FP16 using the same SIMD adder datapath.

However, this execution flow incurs frequent row conflicts between matrix accesses and LUT accesses. Since the tiled matrix and LUT entries reside in different DRAM rows of the same bank, each conversion phase requires the currently open matrix row to be precharged before activating the LUT row. The bank must subsequently reactivate a matrix row to continue operand processing. Therefore, although batching amortizes a LUT-row activation across multiple lookup requests, repeated transitions between operand and LUT rows remain on the critical execution path. To eliminate these repeated matrix-to-LUT row transitions, \name introduces a ping-pong execution scheme that pipelines operand and LUT accesses across the two banks of each even--odd bank pair.

\subsection{Ping-Pong Execution Mechanism} \label{subsubsec:ping_pong}

\begin{figure}
    \centerline{\includegraphics[width=\columnwidth]{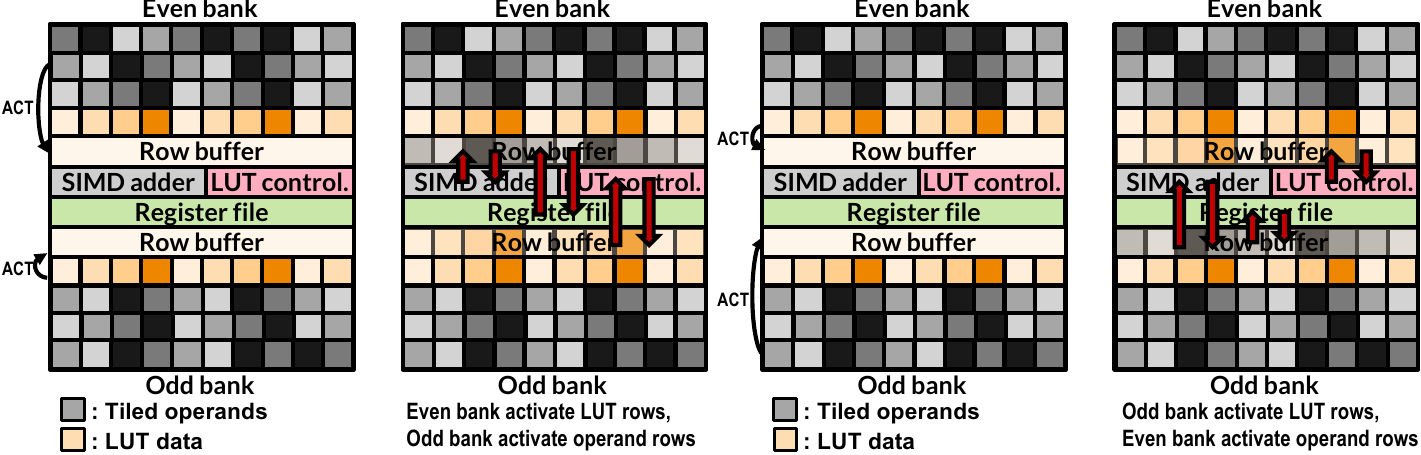}}
    \vspace{-4mm}
    \caption{Proposed ping-pong execution flow.}
    \vspace{-5mm}
    \label{fig:execution_flow_ping_pong}
\end{figure}

Fig.~\ref{fig:execution_flow_ping_pong} illustrates the proposed ping-pong execution flow across an even--odd bank pair sharing a single PU. At any given phase, one bank operates as the \emph{operand bank}, while the other bank operates as the \emph{LUT bank}. Without loss of generality, we assume that the even bank initially acts as the LUT bank and the odd bank as the operand bank.
For each outer-product iteration \(j\), the host broadcasts the vector element \(x_j\) to all PUs, and a 32-byte column access retrieves 16 LNS-encoded matrix elements from the active operand row. The SIMD adder performs the element-wise LNS multiplications as fixed-point additions, and the resulting products are buffered in the register file. Meanwhile, the LUT controller generates the corresponding LUT addresses from the buffered products.

Unlike the basic execution flow, the operand bank does not close its matrix row for number-system conversion. Instead, the generated lookup requests are served by the already-open LUT row in the paired bank. 
Because the LUT row remains active across multiple operand accesses, the controller can issue consecutive LUT column accesses without additional row activations. Even when buffered lookup requests are distributed across different columns, these accesses overlap with SIMD addition in the paired operand bank, hiding most of the lookup latency.
Operand accesses and LUT accesses can therefore be pipelined across the bank pair, hiding a significant portion of the conversion latency while avoiding repeated matrix-to-LUT row conflicts.

After conversion, the FP16 products are accumulated using the same SIMD adder datapath, which is time-multiplexed between fixed-point mode for LNS multiplication and FP16 mode for accumulation. %

Once the operand-bank tiles scheduled for the current phase are processed, the two banks exchange their roles. The odd bank activates its LUT row, while the even bank activates a tiled matrix row, and the same execution flow is repeated for the even-bank operands. Thus, neither bank is permanently dedicated to LUT service, and both banks continue to store matrix tiles and their corresponding output elements.
Although role exchange requires precharge and activation operations, unlike the basic execution scheme, which frequently switches rows between operand-processing and lookup phases, these operations occur only after processing an entire matrix tile. Moreover, the two banks can issue transition commands in parallel. Since each even--odd bank pair already shares one PU, using the non-operand bank for LUT service does not reduce the number of available PUs.

At the final outer-product iteration, the accumulated result is returned to the host in FP16. Returning the FP16 result directly allows subsequent linear operations or LLM layers to consume the output without additional format conversion.

\section{Evaluation}\label{sec:evaluation}

We evaluate \name using an in-house cycle-accurate simulator based on DRAMsim3~\cite{dramsim3}. We assume that the input vector and matrix weights are pre-encoded in LNS before GeMV execution. %
Thus, the reported latency and energy cover the PIM kernel only and exclude host-side FP16-to-LNS16 encoding of the input vector.
The accumulated output is returned in FP16 so that it can be directly consumed by another processing unit for subsequent linear operations or by the host for operations not supported in memory. 
We implement the baseline and proposed logic in RTL and synthesize them using Synopsys Design Compiler with the SAED 14-nm FinFET technology at 250~MHz, the same operating frequency as HBM-PIM. The reported PU area includes the arithmetic units, LUT controller, GRF, SRF, and CRF. For energy estimation, we use the DRAMSim3 HBM2 model, including ACT, PRE, RD, and background energy. We calculate logic energy from post-synthesis power and simulated activity, including the LUT controller and register files.

\subsection{Performance}

\begin{figure}
    \centerline{\includegraphics[width=\columnwidth]{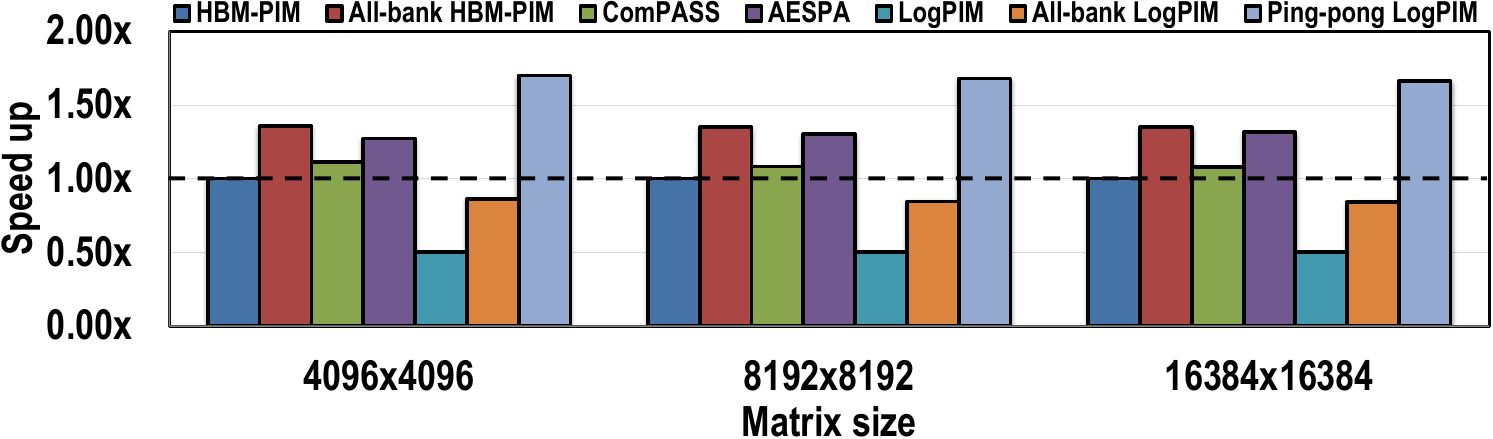}}
    \vspace{-3.5mm}
    \caption{GeMV speedup normalized to HBM-PIM.}
    \vspace{-5mm}
    \label{fig:performance}
\end{figure}

Fig.~\ref{fig:performance} presents the performance comparison. We compare \name with HBM-PIM~\cite{hbm_pim}, an idealized HBM-PIM configuration that activates all banks with a single command, the PIM-only execution scheme of ComPASS~\cite{compass}, and AESPA~\cite{aespa}. For a fair comparison, we evaluate all architectures using the same JEDEC HBM2 timing parameters~\cite{JEDEC_HBM2}, an eight-die HBM2 stack with four PIM-enabled dies, and the register file configuration from HBM-PIM~\cite{hbm_pim}.
Across all evaluated matrix sizes, ping-pong \name achieves average speedups of $1.68\times$, $1.54\times$, and $1.30\times$ over HBM-PIM, ComPASS, and AESPA, respectively. It also achieves average speedups of $1.24\times$ over the idealized all-bank HBM-PIM configuration and $3.36\times$ over the basic \name execution flow. The speedups remain nearly constant across matrix sizes because each architecture executes a fixed command sequence per matrix tile, while the number of tile iterations scales proportionally with matrix size. Applying the same data placement and scheduling to conventional FP16 MAC achieves a 1.69$\times$ average speedup over HBM-PIM, nearly identical to the 1.68$\times$ of ping-pong \name. Thus, the speedup primarily comes from the execution organization, while LNS mainly reduces PU hardware overhead.

The basic execution flow of \name requires approximately $2.00\times$ the execution time of HBM-PIM, primarily because the matrix and LUT reside in the same bank and therefore require repeated row switching. Each lookup precharges the matrix row, activates the LUT row, performs the lookup, and then reactivates the matrix row. Although batching eight LNS products amortizes the LUT activation overhead, it cannot eliminate these row transitions. Separate execution of the even and odd bank sets and command dependencies further prevent matrix accesses from overlapping with LNS-to-FP16 conversion. In contrast, HBM-PIM keeps the matrix row active during MAC operations and accumulates partial sums in local registers.

\subsection{Area and Energy}\label{subsec:area_energy}
The area and energy results are normalized to HBM-PIM. Including the GRF, SRF, and CRF, \name reduces the processing-unit area by 30.1\%. This reduction comes primarily from eliminating the dedicated FP16 multiplier and replacing the conventional FP16 adder with a dual-mode SIMD adder that supports both FP16 and fixed-point addition through shared datapath resources. Although the GRF, SRF, and CRF remain unchanged from HBM-PIM and constitute a substantial fraction of the processing-unit area, \name still achieves a considerable area reduction. As a first-order proportional estimate, this reduces the 50\% capacity loss of a PIM-enabled die to approximately 35\%, increasing the evaluated eight-die stack capacity from 6.0~GB to approximately 6.6~GB; actual recovery depends on physical layout granularity. To support LNS-to-FP16 lookup operations, \name reserves one DRAM row per bank for the conversion LUT, incurring a storage overhead of only one row per bank.

The basic \name execution flow consumes $3.56\times$ the modeled GeMV energy of HBM-PIM. Repeated transitions between the matrix and LUT rows incur additional precharge and activation energy and extend the execution time, thereby increasing background energy. Ping-pong \name avoids these repeated transitions within the inner execution flow by overlapping matrix accesses and LUT operations across paired banks. Consequently, ping-pong \name reduces the modeled GeMV energy by 33.9\% relative to HBM-PIM.

\subsection{Numerical Accuracy}\label{subsec:numerical_accuracy}

\begin{table}%
\centering
\caption{Average GeMV output error compared with the FP32 reference across all evaluated matrix sizes.}
\vspace{-2mm}
\setlength{\tabcolsep}{4pt}
\small
\resizebox{0.85\linewidth}{!}{%
\begin{tabular}{lcccc}
\hline
\textbf{Format} & \textbf{MSE} & \textbf{MAE} & \textbf{MRE} & \textbf{Cosine} \\
\hline
FP16       & $7.32\times10^{-4}$ & $2.04\times10^{-2}$ & $4.18\times10^{-3}$ & $\approx 1.0000$ \\
LNS16 (I6F9)     & $1.05\times10^{-2}$ & $7.73\times10^{-2}$ & $1.30\times10^{-2}$ & $\approx 0.9999$ \\
INT8       & $1.88$              & $1.03$              & $9.49\times10^{-2}$ & $\approx 0.9998$ \\
FP8 (E4M3) & $1.19\times10^{1}$  & $2.60$              & $2.53\times10^{-1}$ & $\approx 0.9993$ \\
\hline
\end{tabular}%
}
\vspace{-6mm}
\label{tab:gemv_precision}
\end{table}

Table~\ref{tab:gemv_precision} compares the GeMV output errors of INT8, FP8, FP16, and LNS16 against the FP32 reference. For each format, the reported error metrics are averaged across the same matrix sizes used in the performance evaluation.
LNS16 uses a 16-bit format consisting of a 1-bit sign, a 6-bit integer, and a 9-bit fraction. LNS16 exhibits substantially lower errors than INT8 and FP8 and maintains a cosine similarity of approximately 0.9999 to FP32, although its absolute errors remain higher than FP16. Thus, the current evaluation demonstrates GeMV-level numerical fidelity rather than end-to-end model accuracy.

\section{Conclusion} \label{sec:conclusion}
We present \name, an LNS-based PIM architecture that reduces the hardware overhead of HBM-based PIM by eliminating the FP16 multipliers and supporting both LNS multiplication and FP16 accumulation with dual-mode SIMD adders. A DRAM-resident LUT and the proposed ping-pong scheme enable efficient LNS conversion while avoiding repeated matrix--LUT row switching. Our evaluation demonstrates that \name achieves low hardware and energy overhead while improving performance and maintaining high GeMV-level numerical fidelity.
As future work, we plan to extend \name to a heterogeneous NPU--PIM system with native LNS support and dynamic, format-adaptive LNS execution.

\bibliographystyle{styles/IEEE/IEEEtran}
\bibliography{bibliography/LOGQ_PIM_ref}

\end{document}